\documentclass{aa}
\usepackage{times,amssymb,amsmath}
\usepackage{epsfig}

\begin{document}

\title{The optical afterglow and host galaxy of GRB~000926 
\thanks{Based on observations made with the Nordic Optical Telescope,
operated on the island of La Palma jointly by Denmark, Finland,
Iceland, Norway, and Sweden.}\fnmsep\thanks{Based on observations made
at the 2.2-m telescope of the German-Spanish Calar Alto Observatory}
\fnmsep\thanks{Based on observations made with the Isaac
Newton Telescope operated on the island of La Palma by the
Isaac Newton Group in the Spanish Observatorio del Roque de los
Muchachos of the Instituto de Astrofisica de Canarias.
}}

\author{
        J.U. Fynbo \inst{1}
        \and J. Gorosabel \inst{2}
        \and T.H. Dall \inst{3,4}
        \and J. Hjorth \inst{5}
        \and H. Pedersen \inst{5}
        \and M.I. Andersen \inst{6}
        \and P. M{\o}ller \inst{1}
        \and S. Holland \inst{11}
        \and I. Smail \inst{7}
        \and N. Kobayashi \inst{8}
	\and E. Rol \inst{9}
        \and P. Vreeswijk \inst{9}
        \and I. Burud \inst{10}
        \and B.L. Jensen \inst{5}
        \and B. Thomsen \inst{4}
        \and A. Henden \inst{12}
        \and F. Vrba \inst{12}
        \and B. Canzian \inst{12}
        \and J.M. Castro Cer\'on \inst{13}
        \and A.J. Castro-Tirado \inst{14,15}
        \and T. Cline \inst{19}
        \and M. Goto \inst{9}
        \and J. Greiner \inst{17}
        \and M.T. Hanski \inst{18}
        \and K. Hurley \inst{16}
        \and N. Lund \inst{2}
        \and T. Pursimo \inst{18}
        \and R. {\O}stensen \inst{20}
        \and J. Solheim \inst{20}
        \and N. Tanvir \inst{21}
	\and H. Terada \inst{22}
}

\institute{
           European Southern Observatory
           Karl-Schwarzschild-Stra{\ss}e 2, D-85748 Garching,
           Germany
           \and
           Danish Space Research Institute,
           Juliane Maries Vej 30, DK--2100 Copenhagen \O, Denmark
           \and
           Nordic Optical Telescope,
           Apartado Postal 474, E-38700 Santa Cruz de La Palma, Spain
           \and
           Institute of Physics and Astronomy, University of \AA rhus,
           DK--8000 {\AA}rhus C, Denmark
           \and
           Astronomical Observatory,
           University of Copenhagen,
           Juliane Maries Vej 30, DK--2100 Copenhagen \O, Denmark
           \and
           Division of Astronomy, 
           P.O. Box 3000, FIN--90014 University of Oulu,
           Finland 
           \and
           Department of Physics, University of Durham, South Road,
           Durham DH1 3LE, UK
           \and
           SUBARU Telescope, National Astronomical Observatory of Japan,
           650 North A'ohoku Place Hilo, 
           Hawaii 96720, U.S.A.
           \and
           University of Amsterdam,
           Kruislaan 403, 1098 SJ Amsterdam, The Netherlands
           \and
           Institut d'Astrophysique et de G\' eophysique, Universit\' e
           de Li\`ege, Avenue de Cointe 5, B--4000 Li\`ege, Belgium
           \and
           Department of Physics, University of Notre Dame,
           Notre Dame, IN 46556-5670, U.S.A.
           \and
	   Universities Space Research Association
           U.S. Naval Observatory,
           Flagstaff Station, Flagstaff, AZ 86002--1149
           \and
	   Real Instituto y Observatorio de la Armada, Secci\'on de
	   Astronom\'{\i}a, 11110 San Fernando--Naval, C\'adiz, Spain
	   \and
	   Instituto de Astrof\'{\i}sica de Andaluc\'{\i}a (IAA-CSIC),
	   P.O. Box 03004, E-18080 Granada, Spain
	   \and
           Laboratorio de Astrof\'{\i}sica Espacial y F\'{\i}sica
           Fundamental
           (LAEFF-INTA), P.O. Box 50727, E-28080, Madrid, Spain
           \and
           University of California, Berkeley,
           Space Sciences Laboratory,
           Berkeley, CA 94720--7450, U.S.A.
           \and
           Astrophysical Institute Potsdam,
           An der Sternwarte 16, 14482 Potsdam, Germany
           \and
           Tuorla Observatory, University of Turku, FIN-21500 Piikki\"o,
           Finland
           \and
           NASA Goddard Space Flight Center,
           Code 661, Greenbelt, MD 20771, U.S.A.
           \and
           Department of Physics, University of Troms{\o}, Troms{\o},
           Norway
           \and
           Department of Physical Sciences,
           University of Hertfordshire, College Lane, Hatfield,
           Herts AL10 9AB, UK
	   \and
	   Department of Physics, Kyoto University, Kitashirakawa
	   Oiwake-cho, Sakyo, Kyoto 606-8502, Japan
           }

\offprints{J.U. Fynbo}
\mail{jfynbo@eso.org}

\date{Received  / Accepted }

\abstract{
We present the discovery of the Optical Transient (OT) of the
long--duration gamma-ray burst GRB~000926. The optical 
transient was detected independently with the Nordic Optical 
Telescope and at Calar Alto 22.2 hours after the burst. At this
time the  magnitude of the transient was R = 19.36. The
transient faded with a decay slope of about $1.7$ during the first two 
days after which the slope increased abruptly (within a few hours)
to about $2.4$. The light-curve started to 
flatten off after about a week indicating the presence of an 
underlying extended object. This object was detected in a deep image 
obtained one month after the GRB at R=23.87$\pm$0.15 and 
consists of several compact knots within about
5 arcsec. One of the knots is spatially coincident with
the position of the OT and hence most likely belongs to the
host galaxy. Higher resolution imaging is needed to 
resolve whether all the compact knots belong to the host
galaxy or to several independent objects. In a separate paper we 
present a discussion of the optical spectrum of the OT, and its
inferred redshift (M{\o}ller et al. in prep.).
\keywords{
cosmology: observations --
gamma rays: bursts}
}

\maketitle

\section{Introduction}

The discoveries of the first $X$-ray afterglow (Costa et al. 1997) and 
OT (van Paradijs et al. 1997) of a gamma-ray burst 
(GRB) have led to a major breakthrough in GRB research. The determination of 
a redshift of 0.835 for \object{GRB~970508} (Metzger et al. 1997), and
the subsequent determination of redshifts of more than a dozen bursts 
with a median redshift of $\sim$1.1, have firmly established their 
cosmological origin (e.g. Kulkarni et al. 2000; Castro-Tirado 2001 and 
references therein).
One of the current goals is to use the OT properties (brightness, position 
within the host galaxy, light-curve shape) as diagnostic tools to study
the environment of GRBs and to possibly shed some light on the nature
of GRB progenitors (e.g. Bloom et al. 2000). 

Another very appealing aspect of GRB research is the use of GRBs
as cosmological probes. GRBs have been suggested as probes of the 
very high redshift universe (Lamb and Reichart 2000; Andersen et al.
2000). An equally
interesting use of GRBs is as an independent way by which to select 
galaxies at cosmological
distances. In some widely accepted scenarios GRBs are related to deaths
of very massive, short-lived stars and furthermore gamma-rays are
not obscured by dust. Hence a sample of GRB host galaxies
may be considered star-formation-selected independent of the
amount of extinction of the rest-frame UV and optical emission. 
However, we do not know if GRBs (as a class) evolve with redshift and 
furthermore
GRBs are of course gamma-ray flux selected. It is nevertheless 
interesting to 
compare GRB-selected galaxies with other samples of galaxies
selected e.g. by rest-frame UV flux (Steidel et al. 1996), Damped Ly-$\alpha$ 
Absorption (DLAs, e.g. Wolfe et al. 1995; Djorgovski et al. 1996; M{\o}ller 
and Warren 1993, 1998; Ellison et al. 2001), Ly-$\alpha$ emission (Hu et al. 
1998; Fynbo et al.
2000a; Kudritzki et al. 2000; Steidel et al. 2000) or sub-mm emission 
(e.g. Ivison et al. 2000). There is currently some
controversy as to whether most of the star-formation at high redshift 
takes place 
in the UV-selected or in the sub-mm selected objects (e.g. Smail et al.
1997; Hughes et al. 1998; Peacock et al. 2000; Adelberger and Steidel 2000;
van der Werf et al. 2000). DLAs trace a very abundant population of
(proto)galaxies at high redshift that could also contribute significantly to
the star-formation density (Fynbo et al. 1999, 2000a). A well understood 
sample of GRB host galaxies may resolve where most of the stars are formed
as a function of redshift.

In this paper we present the detection of the OT and host galaxy
of GRB~000926 and the results of multi-colour optical and
near infrared (IR) photometry.
We then compare the properties of GRB~000926 and its host
galaxy with those of GRB~000301C, which occurred at nearly the same redshift,
with emphasis on the (very different) properties of their 
host galaxies. 

Throughout this paper, we adopt a Hubble constant of H$_0$=65 km s$^{-1}$
Mpc$^{-1}$ and assume $\Omega_m=0.3$ and $\Omega_{\Lambda}=0.7$.

\section{Observations}
\label{observations}

GRB 000926 was detected by three instruments in the Interplanetary
Network (IPN: Ulysses, Konus-WIND, and NEAR), and localized to a
35 arcmin$^2$ error box which was circulated via the GRB Coordinates
Network (GCN)\footnote{
{\tt \small http://gcn.gsfc.nasa.gov/gcn/}} 
20.3 hours after the
burst (Hurley 2000).  The Earth-crossing time for the burst was
September 26.9927 UT. As observed by Ulysses, it had a duration of
approximately 25 seconds (placing it in the "long duration" burst
category), a revised 25-100 keV fluence of 6.2$\times$10$^{-6}$ erg 
cm$^{-2}$, and a revised peak flux over 0.25 s of 8.6$\times$10$^{-7}$ 
erg cm$^{-2}$ s.

\begin{figure*}
\begin{center}
\epsfig{file=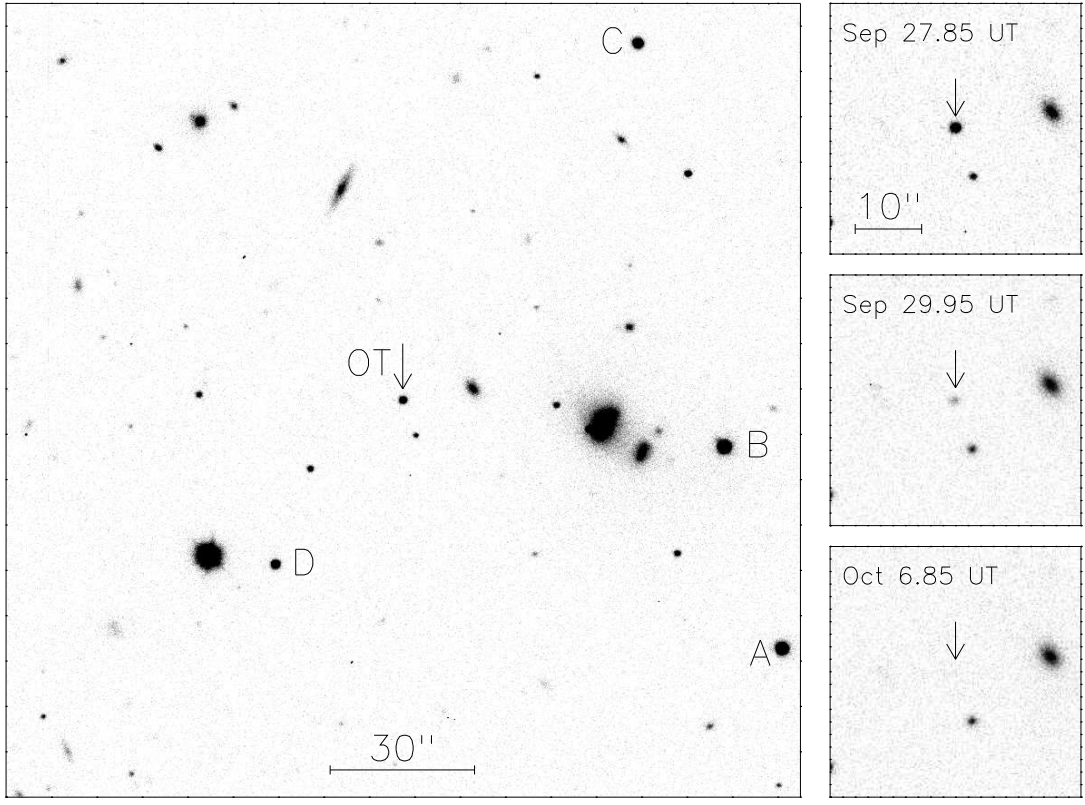,width=15cm}
\caption{{\it Left panel\/}: An R--band finding-chart for GRB~000926 
centred on the position of the OT (marked with an arrow). Also marked 
are the four secondary calibration stars A--D used to transform the
relative PSF-photometry onto the standard system. The photometric
properties of the stars are given in Table~\ref{compphot}.
{\it Right panels\/}: Three smaller R--band images centred on the OT at 
three different epochs showing the decline of the OT.
}
\label{OT}
\end{center}
\end{figure*}

The error-box of \object{GRB~000926} (Hurley 2000) 
was observed in the R--band with the 2.56-m Nordic Optical Telescope 
(NOT) and the Calar Alto (CA) 2.2-m telescope on 2000 September 27.85 
UT (20.64 hours after the burst). Comparing with red 
Palomar Optical Sky Survey~II exposures an OT was found in the
error box (Gorosabel et al.
2000; Dall et al. 2000). Spectra were obtained at the NOT on
both September 27 and September 28; these spectra revealed a strong 
metal absorption system at a redshift of z$=2.0375\pm0.0007$, which in
all likelihood is due to gas in the GRB host
galaxy (M{\o}ller et al. in prep). The OT was observed during 
the following weeks at the NOT, CA and also with the Isaac Newton
Telescope (INT), the 1.0 m telescope at the US Naval Observatory at
Flagstaff Station (USNOFS), and in the IR with the 8.2-m Subaru telescope
and the 3.8-m United Kingdom Infrared Telescope (UKIRT).
The logs of optical and IR observations are given 
in Table~\ref{obslog} and Table~\ref{IRlog}. A finding-chart and 
R--band images of the
OT at three different epochs are shown in Fig.~\ref{OT}.

%=====================Begin Figure OT===========================
%=====================End Figure 1=============================

%=====================Begin Table 2==============================
\begin{table*}
\begin{center}
\caption{UBVRI magnitudes for the four secondary calibration stars A--D}
\begin{tabular}{@{}lcccccc}
star/filter  &    U      &   B    &    V     &    R     &    I   \\
\hline
A            & 17.928$\pm$0.069 & 18.499$\pm$0.021 & 17.635$\pm$0.010 & 17.025$\pm$0.021 & 16.553$\pm$0.028 \\
B            & 17.728$\pm$0.118 & 18.662$\pm$0.025 & 17.508$\pm$0.010 & 16.811$\pm$0.022 & 16.265$\pm$0.030 \\
C            & 19.174$\pm$0.086 & 19.047$\pm$0.036 & 18.449$\pm$0.016 & 18.026$\pm$0.028 & 17.655$\pm$0.038 \\
D            &        $-$       & 20.822$\pm$0.141 & 19.459$\pm$0.045 & 18.529$\pm$0.065 & 17.694$\pm$0.069 \\
\hline
\label{compphot}
\end{tabular}
\end{center}
\end{table*}
%=====================End Table 2==============================
%=====================Begin Table 1==============================
\begin{table}[t]
\begin{center}
\caption{The journal of NOT, CA, INT and USNOFS observations and the 
results of the photometry.}
\begin{scriptsize}
\begin{tabular}{@{}llccccc}
UT           & Obs. & mag & Seeing    & Exposure time \\
             &      &     & (arcsec)  & (sec) \\
\hline
{\it U--band}$^a$ \\
Sep 27.9484  & NOT &  20.437$\pm$0.047  & 1.35 & 1000 \\   
Sep 28.8291  & NOT &  21.732$\pm$0.224  & 1.29 &  450 \\
Sep 29.8493  & NOT &  22.816$\pm$0.131  & 1.16 &  900 \\
{\it B--band}$^a$ \\
Sep 27.8881  & NOT &  20.432$\pm$0.034  & 1.00 &  100 \\
Sep 27.9673  & CA & 20.733$\pm$0.034    & 2.42 & 1200 \\
Sep 27.9894  & CA & 20.727$\pm$0.038    & 2.41 & 1200 \\
Sep 28.0057  & CA & 20.796$\pm$0.039    & 2.59 & 1200 \\
Sep 29.8346  & NOT &  22.890$\pm$0.122  & 1.19 &  500 \\
Sep 29.8688  & NOT &  22.856$\pm$0.082  & 1.20 &  600 \\
Sep 30.8688  & NOT &  23.489$\pm$0.097  & 1.20 & 1200 \\
{\it V--band}$^{a,b}$ \\
Sep 27.906   & INT & 19.906$\pm$0.0135  & 1.0  &  300 \\
Sep 27.910   & INT & 19.933$\pm$0.0135  & 1.0  &  300 \\
Sep 27.915   & INT & 19.936$\pm$0.0141  & 1.0  &  300 \\
Sep 27.9237  & CA  & 19.910$\pm$0.025   & 2.22 &  900 \\
Sep 29.9557  & NOT &  22.324$\pm$0.121  & 1.29 &  600 \\
Sep 30.839   & INT & 22.987$\pm$0.096   & 1.0  &  600 \\
{\it R--band}$^{a,b}$ \\
Sep 27.8547  & NOT &  19.326$\pm$0.015 & 0.83 & 300  \\ 
Sep 27.8594  & NOT &  19.343$\pm$0.011 & 0.84 & 300  \\
Sep 27.8639  & NOT &  19.322$\pm$0.011 & 0.86 & 300  \\
Sep 27.8651  & CA  &  19.329$\pm$0.034 & 1.44 & 900  \\
Sep 27.8684  & NOT &  19.342$\pm$0.008 & 0.82 & 300  \\
Sep 27.8729  & NOT &  19.349$\pm$0.011 & 0.87 & 300  \\
Sep 27.8774  & NOT &  19.366$\pm$0.010 & 0.90 & 300  \\
Sep 27.8820  & NOT &  19.349$\pm$0.010 & 0.91 & 300  \\
Sep 27.892   & INT &  19.394$\pm$0.022 & 1.0  & 300  \\
Sep 27.896   & INT &  19.411$\pm$0.022 & 1.0  & 300  \\
Sep 27.901   & INT &  19.421$\pm$0.023 & 1.0  & 300  \\
Sep 27.9109  & CA  &  19.452$\pm$0.020 & 2.01 & 900  \\
Sep 27.9658  & NOT &  19.538$\pm$0.011 & 1.06 & 300  \\
Sep 27.9705  & NOT &  19.526$\pm$0.014 & 1.09 & 300  \\
Sep 27.9750  & NOT &  19.567$\pm$0.015 & 1.01 & 300  \\
Sep 27.9795  & NOT &  19.542$\pm$0.017 & 1.02 & 300  \\
Sep 27.9840  & NOT &  19.553$\pm$0.023 & 1.00 & 300  \\
Sep 28.097   & USNOFS & 19.71$\pm$0.03 & 2.0  & 3$\times$600 \\
Sep 28.8356  & NOT &  20.723$\pm$0.025 & 1.06 & 300  \\
Sep 28.8911  & NOT &  20.774$\pm$0.036 & 1.05 & 500  \\
Sep 28.9595  & NOT &  20.875$\pm$0.036 & 1.19 & 600  \\
Sep 29.110   & USNOFS & 20.94$\pm$0.04 & 1.8  & 3$\times$600 \\
Sep 29.138   & USNOFS & 20.95$\pm$0.04 & 1.9  & 4$\times$600 \\
Sep 29.170   & USNOFS & 21.08$\pm$0.05 & 2.0  & 4$\times$600 \\
Sep 29.8414  & NOT &  21.740$\pm$0.043 & 0.86 & 600  \\
Sep 29.9478  & NOT &  21.897$\pm$0.083 & 1.17 & 600  \\
Sep 30.83    & NOT &  22.494$\pm$0.070 & 0.90 & 2400 \\
Oct  2.84    & NOT &  23.514$\pm$0.153 & 1.14 & 1500 \\
Oct  6.85    & NOT &        $-$        & 1.29 & 2700 \\
Oct 27.84    & NOT &        $-$        & 1.10 & 4700 \\
Nov  3.84    & NOT &        $-$        & 1.26 & 2400 \\
Nov  4.84    & NOT &        $-$        & 1.12 & 3600 \\
{\it I--band}$^{a,b}$ \\
Sep 27.919   & INT &  18.959$\pm$0.029 & 1.0  &  300 \\
Sep 27.924   & INT &  18.992$\pm$0.029 & 1.0  &  300 \\
Sep 27.929   & INT &  19.007$\pm$0.029 & 1.0  &  300 \\
Sep 27.9370  & CA  &  18.861$\pm$0.032 & 2.22 &  900 \\
Sep 27.9433  & NOT &  19.014$\pm$0.014 & 0.72 &  600 \\
Sep 27.9510  & CA  &  18.890$\pm$0.022 & 2.16 &  900 \\
Sep 29.8608  & NOT &  21.218$\pm$0.051 & 0.79 &  600 \\
Sep 30.839   & INT &  21.956$\pm$0.029 & 1.0  &  600 \\
\hline
\label{obslog}
\end{tabular}
\end{scriptsize}
\end{center}
\begin{footnotesize}
\vskip -0.5cm

\noindent
$^a$ Only for the NOT R--band measurements have we subtracted a galaxy 
image before doing the PSF photometry.

\noindent
$^b$ The following filters were used: NOT -- all Bessel; CA -- all Johnson;
INT: Sloan i, Harris R \& V; USNOFS : Bessel R.
\end{footnotesize}
\end{table}
%=====================End Table 1==============================

Radio and X-ray afterglow measurements have been reported by
Frail et al. (2000) and by Piro et al. (2000). Optical afterglow
measurements have also been reported by Price et al. (2001).
This paper includes preliminary reduced NOT and CA observations
from Hjorth et al. (2000), Fynbo et al. (2000b) and Gorosabel et
al. (2000) for which
the finally reduced data appear in this paper.

\section{Celestial Position}
By measuring the position of the OT relative to 80 stars in the
USNO-A2.0 catalog we found the celestial coordinates of the OT to be
RA(J2000) = 17:04:09.68, Dec(J2000) = +51:47:10.5 with an internal
error of about 0.1 arcsec and a systematic error of about 
0.25 arcsec (Deutsch 1999).

\section{Photometry}
\subsection{R--band light-curve of the OT}
\label{lc}
In comparison with \object{GRB~000301C} (Jensen et al. 2001)
GRB~000926 was found 
to be hosted by a relatively luminous host galaxy (see Sect.~\ref{host}). 
When measuring the magnitude of the OT directly on the images with Point 
Spread Function (PSF) photometry there will be a contribution from 
the underlying host galaxy. The relative strength of this contribution 
depends on the seeing and is therefore a source of systematic errors if not
corrected for. We therefore subtracted an aligned and scaled
image of the host galaxy from each of the individual R--band images
from the NOT listed in Table~\ref{obslog}.  This image of the host
galaxy was obtained about 30 days after the burst when the  
magnitude of the OT, based on an extrapolation of the light-curve,
as well as a possible underlying supernova would
be negligible (see Sect.~\ref{host}).
We then measured the magnitude of the OT using DAOPHOT-II 
(Stetson 1987, 1997). The OT was last detected in the image obtained on 
October 2. The photometry was transformed to the standard system using
photometric observations of the four secondary reference stars A--D
marked in Fig.~\ref{OT} obtained at the U.S. Naval Observatory
Flagstaff Station. The calibrated UBVRI magnitudes of the stars A--D are 
given in Table~\ref{compphot}.

We first fitted a broken power-law 
\begin{equation}
\label{EQUATION:broken_power_law}
f_{\nu}(t) = \left \{
        \begin{array}{lll}
                f_{\nu}(t_b) {\left(\frac{t}{t_b}\right)}^{-\alpha_1}, &
                \mathrm{if}  &  t \le t_b \\
                f_{\nu}(t_b) {\left(\frac{t}{t_b}\right)}^{-\alpha_2}, &
                \mathrm{if}  &  t \ge t_b,
        \end{array}
             \right.
\nonumber
\end{equation}
to the data, which provided a very good fit. The parameters of
the fit are given in Table~\ref{fits}.
We then followed Beuermann et al. (1999) and fitted an empirical function 
of the form
\[f_{\nu}(t) = (f_1(t)^{-n} + f_2(t)^{-n})^{-1/n},\]
where $f_i(t) = k_i~t^{-\alpha_i}$ and $t$ is the time since the
GRB measured in days. For large values of $n$ this function approaches
the broken power-law. We performed this fit both keeping $n$ fixed
at 1 similar to Stanek et al. (1999) and with $n$ as a free parameter. 
The results of these fits are also given in Table~\ref{fits}. 

The $\chi^2$ 
per degree of freedom is smallest for the broken power-law fit.
The data favor a very large value of $n$ as the $\chi^2$ is monotonically 
decreasing as $n$ is increased even for $n > 100$. The 2$\sigma$ lower limit 
on $n$ is $7$ in the sense that the difference between $\chi^2$ for a fit with 
$n = 7$ and the broken power-law is 2. This indicates that the break in the 
light-curve is very abrupt (within a few hours). 

\begin{figure}
\begin{center}
\epsfig{file=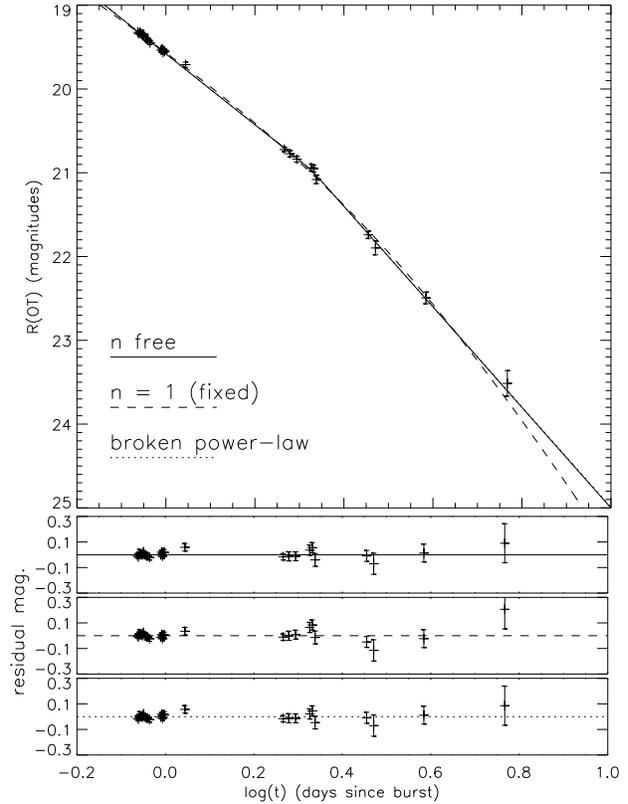,width=8.5cm}
\caption{The R--band light-curve of the OT. The upper panel show
the R--band measurements together with the three light-curve fits.
The three lower panels shows the deviations around the fits for,
from top to bottom, the broken power-law, the $n$ fixed and the
$n$ free fits.}
\label{lightcurve}
\end{center}
\end{figure}

In conclusion, the data 
are best fit by a sharp break around t=2.12$\pm$0.09 days after the burst.
In Fig.~\ref{lightcurve} we show the R--band 
light-curve of the OT together with the three fits and the residuals 
around the fits.

%=====================Begin Figure LC==========================
%=====================End Figure LC============================

%=====================Begin Table 3==============================
\begin{table}[t]
\begin{center}
\caption{Results of fits to the R--band light-curve. There were 24 degrees
of freedom in the fits.}
\begin{tabular}{@{}lcccccc}
Fit            &  $\alpha_1$  &  $\alpha_2$  &  other  &  $\chi^2$/dof \\
\hline   
b. p.-l.       &  1.69$\pm$0.02 & 2.39$\pm$0.09 & $t_{b}=2.12\pm0.09$ & 1.000 \\
n=1            &  1.40$\pm$0.13 & 3.36$\pm$0.41 & $n=1$ (fixed) & 1.263 \\
n free         &  1.69$\pm$0.02 & 2.39$\pm$0.09 & $n>7$ (2$\sigma$) & 1.084 \\
\hline
\label{fits}
\end{tabular}
\end{center}
\end{table}
%=====================End Table 3==============================

\subsection{Other optical bands}
 The OT was also imaged in the U,B,V and I bands during
the first four nights after the detection at the NOT and CA. 

To precisely determine the broad band colours of the OT we used the 
UBV and I band observations obtained at NOT and INT. The CA points were excluded
since Johnson R and I filters are significantly different from Bessel, 
Cousins and
Harris. We determined the colours as the offset of the broken power-law fit
to the R--band light-curve that minimized the $\chi^2$ of the fit. 
Due to the lower error bars the magnitudes obtained on Sep 27 have
the largest weight in the fits. Therefore we do not expect a 
large systematic uncertainty due to emission from the underlying
host galaxy. In order to minimize the effect of the host galaxy
only points obtained earlier than and including September 29 were used
in the fits.  The 1$\sigma$ errors on the
colours were determined as the colours that increased the value
of $\chi^2$ by 1, but the true uncertainty including calibration and
systematic errors is most likely somewhat
larger ($\sim$5\%). For all filters U,B,V and I the fits were 
consistent with the (offset) broken power-law fit, which shows that
the data within the errors (few percent) are consistent with an 
achromatic optical afterglow. The results are given in 
Table~\ref{colours}. 

%=====================Begin Table 4==============================
\begin{table}
\begin{center}
\caption{Colours of the OT of GRB~000926 and GRB~000301C. The
colours of the OT of GRB~000301C are from Jensen et al. (2001).}
\begin{tabular}{@{}lcccccc}
Colour & GRB~000926  & $\chi^2$/dof &  & GRB~000301C  \\ 
\hline
U-R  & 0.96$\pm$0.04 & 0.59     &   & 0.40$\pm$0.05 \\  
B-R  & 1.06$\pm$0.03 & 0.53     &   & 0.91$\pm$0.03 \\ 
V-R  & 0.50$\pm$0.01 & 0.34     &   & 0.44$\pm$0.05 \\  
R-I  & 0.48$\pm$0.01 & 0.48     &   & 0.49$\pm$0.04 \\
R-K  & 3.35$\pm$0.07 & 1.87     &   & 2.98$\pm$0.08 \\
\hline
\label{colours}
\end{tabular}
\end{center}
\end{table}
%=====================End Table 4==============================

\subsection{Infrared photometry}
The afterglow was observed in the K'--band on September 29, 2000
with the IRCS instrument on the 8.2-m Subaru telescope in a seeing 
of about 0.7 arcsec and a total integration time of 1800sec. No
standard star observations were obtained on the same night,
but fortunately there is a bright, unsaturated 2MASS source in
the field which we could use for calibration. 
We measured the magnitude of the OT 
and of the 2MASS source in a 2 arcsec circular aperture. The formal 
photometric error-bar is less than a percent, but we estimate 
conservatively that the uncertainty in the colour-transformation 
for the IRCS instrument amounts to 0.10 mag.    

The afterglow was also observed in the J, H and K bands with the UFTI
imager on UKIRT
on September 30. The final frames were accumulated in 26, 24 and
9 dithered exposures of 60\,s duration for J, H and K respectively 
resulting in a total on-source integration time of 1.56\,ks, 1.44ks
and 540\,s, all in photometric conditions. Employing standard
procedures these frames were reduced, combined and calibrated
using observations of UKIRT faint standards bracketing the
science exposures.  The final frames have only a 
(for this instrument) modest seeing of
0.55--0.60$''$ FWHM due to the relatively high airmass of the
observations, 1.5--2, but clearly detect the OT in all three
passbands. The magnitude of the OT was again measured using
aperture photometry. The results of the IR photometry is presented
in Table~\ref{IRlog}. 

Using the standard star calibrated UKIRT K-band observations we
confirmed from faint objects visible in both the UKIRT and
SUBARU images that the calibration of the SUBARU images is
consistent with that of the UKIRT K-band observations.

%=====================Begin Table 4==============================
\begin{table}
\begin{center}
\caption{Log of IR observations}
\begin{tabular}{@{}lllcccc}
UT (Sep)     & filter/Obs. & mag & Seeing    & Exp. time \\
             &      &     & (arcsec)  & (sec) \\
\hline
30.276   & J/UKIRT & 20.83$\pm$0.15 & 0.6 &  1560 \\
30.250   & H/UKIRT & 19.46$\pm$0.10 & 0.6 &  1440 \\
29.24    & K'/SUBARU & 17.86$\pm$0.10 &  0.7  &  1800  \\
30.301   & K/UKIRT  & 18.66$\pm$0.11 &  0.6  &  540  \\
\hline
\label{IRlog}
\end{tabular}
\end{center}
\end{table}
%=====================End Table 4==============================

\section{Spectral energy distribution of the afterglow}

Also  shown in Table~\ref{colours} are the  colours (in  the same bands) of
the OT of GRB~000301C, which was also discovered at the NOT and had a
very similar redshift as GRB~000926 (Jensen et al. 2001; M\o ller et al.
in prep). 
As seen, the OTs of GRB~000926 and GRB~000301C had very similar colours in
the  optical red bands, whereas in the blue bands and in R$-$K the OT of 
GRB~000926 was
significantly redder than that  of GRB~000301C.  In  order to test  whether
this difference is intrinsic to the bursts or caused by a larger extinction
along the line  of sight to  GRB~000926 we follow Jensen et al. (2001)
and constrain the extinction by fitting different extinction laws to the
SED.

\begin{figure*}[t]
\begin{center}
  \epsfig{file=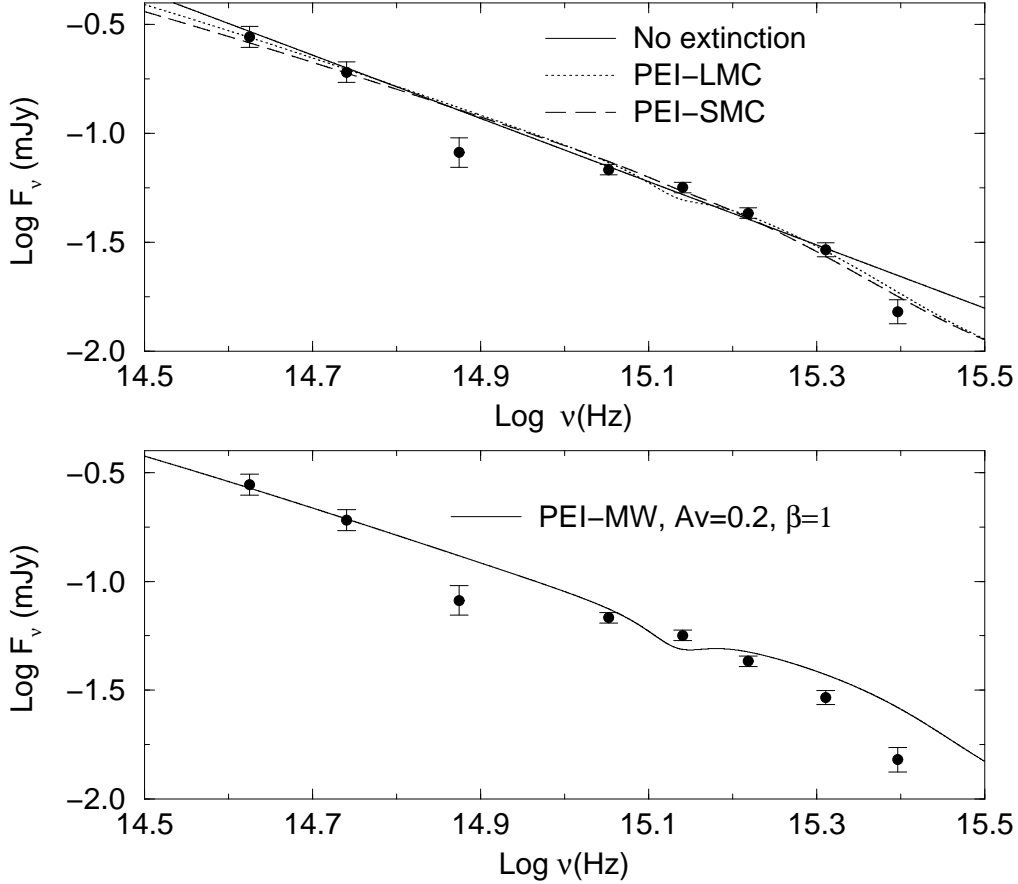,width=15cm,angle=0,clip=}
\caption{The observed specific flux F$_{\nu}$ vs. the rest-frame
frequency $\nu$ normalised to Sep 27.9 UT. The spectral 
energy distribution of the afterglow from K to U is clearly 
curved. In the upper panel we show that this curvature can
be well fitted by an underlying $\beta=1$ power-law SED with
a SMC or LMC extinction law with A$_{\rm V}$=0.2. In the lower 
panel we show that the MW extinction law is inconsistent with
the data.}
\label{sed}
\end{center}
\end{figure*}

To construct the SED we first used the colours given in Table~\ref{colours} 
for the observed U to I bands (normalised to Sep 27.9 UT). 
The J, H and K-observations were obtained on Sep 30.3 where
the host galaxy possibly contributed significantly to the flux.  
In order to estimate the effect of the host galaxy
we used the SEDs for galaxies at redshifts z=2--3 given by
Dickinson (2000, their Fig. 2). By normalising these galaxy SEDs
to the observed R(AB)=24.04$\pm$0.15 for the host galaxy (see
Sect.~6 below) we derived magnitudes for the host galaxy
which translate into estimated corrections at Sep 30.3 UT of
$\Delta$J=+0.14$\pm$0.08, $\Delta$H=+0.07$\pm$0.06 and 
$\Delta$K=+0.05$\pm$0.04. The JHK magnitudes were then shifted to
Sep 27.9 UT using the broken power-law fit to the light-curve given 
in Table~\ref{fits} (assuming that the burst evolved achromatically).
After this the UBVRIJHK magnitudes
were corrected for foreground extinction, using a value of 
E(B$-$V)=0.023 from Schlegel et al. (1998), and transformed to
the AB system. For the optical bands we used the transformations 
given by Fukugita et al. (1995): I(AB) = I+0.43, R(AB) = R+0.17, 
V(AB) = V-0.02, B(AB) = B-0.14, and U(AB) = U+0.69. We assigned 
uncertainties of 0.05 mag to the BVR and I AB magnitudes 
as an estimate of the uncertainty in the transformation. For
U band we assigned an uncertainty of 0.10 mag to the AB magnitude 
since this band is more difficult to calibrate (Bessel 1990; Fynbo
et al. 1999, 2000c). For the IR bands we used the
transformations given in Allen (2000): K(AB) = K+1.86, H(AB) = H+1.35,
and J(AB) = J+0.87. We then calculated the
specific flux using $F_{\nu} = 10^{-0.4\times(AB+48.60)}$.
Finally, the wavelengths corresponding to our
UBVRIJHK measurements were blueshifted to the GRB rest frame.  As it can be
seen in Fig.~\ref{sed} the spectral energy distribution is clearly
bending from the U to the K--band. This bend can be naturally explained by
the presence of intrinsic extinction at z=2.037. The J-point is falling
significantly below the trend of all the other points. The reason for
this is not understood, but we have decided not to include this point in 
the analysis. Including the point does not change any of the conclusions,
but it increases the $\chi^2$ of the fits. 

We next fitted the function F$_{\nu} \propto \nu^{-\beta} \times 
10^{(-0.4 {\rm A}_{\nu})}$ to the SED. Here, $\beta$ is the spectral
index and A$_{\nu}$ is the extinction in  magnitudes at frequency $\nu$.
We have considered the three extinction laws (A$_{\nu}$ as a function
of $\nu$) given by  Pei (1992), i.e. for  the  Milky-Way (MW), Large 
Magellanic Cloud (LMC) and the Small Magellanic Cloud (SMC). In the 
three cases the dependence of the extinction with $\nu$ have been  
parameterized in terms of (restframe) A$_{\rm V}$. Thus, our fits allow 
us to  determine $\beta$ and A$_{\rm V}$ simultaneously. Finally,
we also considered the no-extinction case where F$_{\nu}$ was fitted
by a straight line in log-log space.

The parameters of the fits are shown in Table~\ref{sedtab}.  
For the no-extinction case we find a value of $\beta$ consistent
with that of Price et al. 2001.
As for GRB~000301C the best fit was achieved for a 
SMC extinction law. We derive a modest  extinction of 
A$_{\rm V}=0.18{\pm}0.06$ (restframe $V$) and a spectral index 
$\beta=1.00{\pm}0.18$. For
GRB~000301C Jensen et al. (2001) found $\beta = 0.70\pm0.09$. Therefore,
GRB000926 was indeed intrinsically redder than GRB~000301C. 

In the upper panel of Fig.~\ref{sed} we show the fits using the LMC and SMC
extinction laws and the no-extinction case.
For the redshift of GRB~000926 (as for that of GRB~000301C) the interstellar
extinction bump  at 2175 \AA \ is shifted into the R-band. This absorption 
bump is very prominent for the MW, moderate for  the LMC and 
almost nonexistent for the SMC  extinction curve. Thus, for a chemically
rich environment, like  the MW, we should  expect a prominent extinction
bump  at 2175 \AA \ (near the observed R-band).
 The data points in Fig.~\ref{sed} shows that there is no strong 
absorption bump near the 
R-band, which makes the fit for the MW (see  Table~\ref{sedtab})  
inconsistent with the data. In fact, the best MW fit implies
a (unphysical) negative extinction. To illustrate the problem with
the MW extinction curve we have in the lower panel of Fig.~\ref{sed}
plotted a $\beta=1$ power-law SED extincted by a A$_{\rm V}$=0.2 MW
extinction curve. As seen, the shape of this extinction curve is
incompatible with the data. In the Milky Way the extinction curve 
can be different mainly for stars located in star-forming regions
(Baade and Minkowski 1937; Whittet 1992) in the sense that the shape 
of the bump  at 2175 \AA \ is different and more importantly the 
curve is almost flat in the rest-frame UV at $\log(\nu)$ $>$ 15.1.
This is where the curvature is most pronounced in Fig.~\ref{sed}
and therefore such an extinction curve is also not compatible with
the data (see also Price et al. 2001)

In conclusion, as in the
case of GRB000301C, the  SED supports a  scenario of a host in an early
stage of chemical enrichment.

%==================== Begin Table sed =======================

\begin{table}[t]
\begin{center}
\caption{The fits to the spectral energy distribution of GRB000926}
\begin{tabular}{lccr}
   & $\chi^2/{\rm dof}$ & $\beta$ & A$_{\rm V}$  \ \ \ \ \ \ \ \\
\hline
 No extinction  & 3.20 & $-$1.42 ${\pm}$ 0.06 & 0 \ \ \ \ \ \ \ \ \  \\
\hline
Pei (1992),  MW &      &                   &  $<$0 \ \ \ \ \ \ \ \ \ \\
Pei (1992), LMC & 2.61 & 0.98 ${\pm}$ 0.23 &  0.27 ${\pm}$ 0.12\\
Pei (1992), SMC & 1.71 & 1.00 ${\pm}$ 0.18 &  0.18 ${\pm}$ 0.06\\
\hline
\label{sedtab}
\end{tabular}
\end{center}
\end{table}

%==================== End Table sed =========================

%=====================Begin Figure sed========================
%=====================End Figure sed==========================

\section{The host galaxy}
\label{host}
   In the images obtained on October 6 the emission at the position of the OT 
appeared elongated although at low signal-to-noise ratio. We therefore obtained 
further deep imaging on October 27 and November 3 and 4. A total of 3 hours of 
R-band imaging was obtained under dark/grey sky conditions and a seeing varying
between 1.0 and 1.3 arcsec. We subsequently combined these images using the 
code described in M{\o}ller and Warren (1993). The FWHM of point sources in the
combined image was 1.18 arcsec. In the combined image an extended object 
directly underneath the position of the OT was clearly detected. A contour 
image showing this extended emission is shown as Fig.~\ref{host}. 
Although the seeing was not optimal, the object is resolved into several
compact knots covering a region of total extension of about 5 arcsec. This
is more clearly visible in images of the host galaxy + OT obtained with
the Hubble Space Telescope presented by Price et al. (2001). In order 
to determine the position of the OT relative to these compact knots we measured
the position of the OT relative to point-sources in the vicinity in the first 
image obtained on September 27 and in the combined image. The position of the 
OT in the combined image could in this way be determined with a  
conservatively estimated uncertainty of 0.02 arcsec (10\% of a pixel). The 
position of the OT is within the errors 
spatially coincident with one of the knots of the extended emission. As
the OT,
based on an extrapolation of the light-curve, is expected to be much fainter 
than this knot (by several magnitudes) we conclude that at least this 
particular knot is related to the host galaxy of GRB~000926. An underlying 
supernova (SN) even if similar to SN1998bw would also be much fainter than the
observed knot, especially since the R--band corresponds to the rest-frame UV
around 2100\AA \ where SNe are intrinsically faint. Whereas it is possible that 
all the compact knots are either part of the host galaxy or several  
galaxy sub-clumps in the process of merging, we cannot exclude that some of the 
emission is from objects lying at other distances along the line of sight
(although there are no intervening absorption systems in the spectrum of the
OT, M\o ller et al. in prep).
If all the emission is from the host galaxy of 
GRB~000926 then the extension of the galaxy would be similar to the
Ly-$\alpha$ emitting region of the galaxy S4 at z=1.93 towards Q0151+048A 
(Fynbo et al. 1999).

The total magnitude of the extended emission in a circular aperture with 
diameter 
4.7 arcsec is R=23.87$\pm$0.15. Assuming that all this emission comes from
the host galaxy we can get an estimate of the star-formation-rate (SFR) of
the galaxy. The restframe UV continuum in the range 1500\AA--2800\AA \
can be used as a SFR estimator if one assume that the star-formation is 
continuous over a time scale of more than 10$^8$ years. Kennicutt (1998) 
provides the relation 
\[
SFR(M_{\sun} yr^{-1}) = 1.4 \times 10^{-28} \times L_{\nu},
\]
where $L_{\nu}$ is the luminosity in the 1500\AA--2800\AA \ range
measured in erg s$^{-1}$ Hz$^{-1}$. The observed R--band corresponds to 
the rest-frame UV continuum around 2100\AA, which falls well within
this range. To derive $L_{\nu}$ we first converted our R--band 
magnitude to R(AB) using R(AB)=R+0.17 (Fukugita et al. 1995).
Then we used the definition of the AB magnitude to derive the observed
flux ($F_{\nu} = 10^{-0.4\times(R(AB)+48.6)}$) and finally the 
luminosity distance in our assumed cosmology 
(d$_{lum}$=5.28$\times$10$^{28}$cm) to derive $L_{\nu}$ :
\begin{eqnarray}
L_{\nu} &=& F_{\nu} \times 4 \pi d_{lum}^2/(1+z) {}
\nonumber\\
&=& {} (1.02\pm0.15) \times 10^{29}\: erg s^{-1} Hz^{-1} {},
\nonumber
\end{eqnarray}
where the factor $(1+z)^{-1}$ corrects for the fact that $L_{\nu}$ is
a specific luminosity (not a bolometric luminosity). Using the relation 
of Kennicutt (1998) we find a SFR of 14 M$_{\sun}$ yr$^{-1}$.
If the extinction derived in Sect.~5 is valid
for the galaxy as a whole then we estimate an extinction in the observed 
R-band of 0.56$\pm$0.20 mag, which means that the SFR should be increased
by a factor of $\sim1.7$.
This SFR is a high compared to that of the Milky Way, but it falls in
the low end of the range of SFRs of Lyman-Break galaxies
(Pettini et al. 1998).

%=====================Begin Figure host========================
\begin{figure}
\begin{center}
\epsfig{file=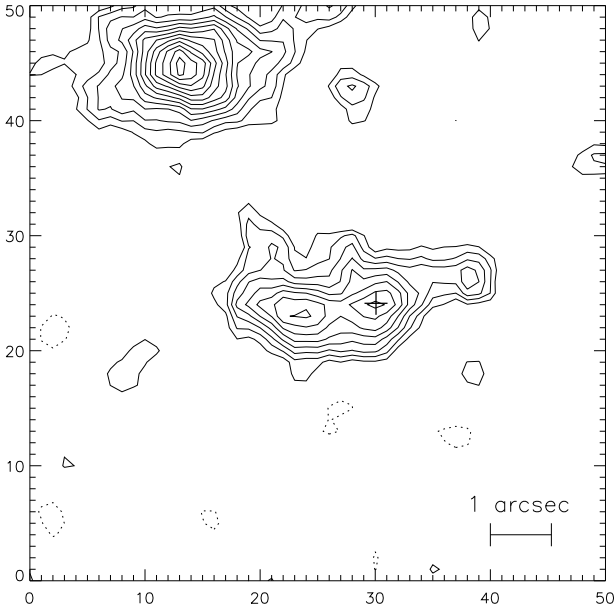,width=8.5cm}
\caption{A contour plot of a 9.5$\times$9.5 arcsec$^2$ region
centred on the probable host galaxy of
GRB~000926. The image has been smoothed with a Gaussian filter
with a width of 2 pixels (0.39 arcsec). The contour levels are
-2, 2, 3.5, 5, 6.5, ... $\times$ 1$\sigma$ of the sky-noise, with
the dotted contours being negative. East is left and north is 
up. The cross marks the position of the OT. The error on the 
position of the OT is 0.10 pixels corresponding to 0.02 arcsec.}
\label{hostcont}
\end{center}
\end{figure}
%=====================End Figure host==========================

\section{Discussion}

\subsection{Interpretation of the light-curve}
  The temporal behavior of the optical afterglow of GRB~000926 is a 
clear and unambiguous example of a broken power-law decay. The
fits described in Sect.~\ref{lc} show that the break
occurred abruptly. The late time decay slope of 
$\alpha_2$=2.39$\pm$0.09 is very similar to the late time
decay slopes of other well studied broken or fast decaying 
light-curves (see e.g. Andersen et al. 2000, their Fig.~4 and 
Table~4). This striking uniformity suggest a common physical 
scenario for the late stage of the decays, which most likely
is a common value of the index of the electron energy 
distribution (see also Sari et al. 1999; Freedman and Waxman
2001). 

  The increase $\Delta\alpha = \alpha_2 - \alpha_1$ 
from the early to the late time decay slope is different for 
different physical models for GRB afterglows. For 
GRB~000926 we find $\Delta\alpha = 0.70\pm0.09$ from the broken 
power-law fit. 
This measurement we compare with different models predicting 
broken light-curves: {\it i)\/} If the frequency separating fast 
cooling and slow cooling electrons moves through the optical part of 
the electromagnetic spectrum at
$t_b$, the resulting light curve would steepen by $\Delta \alpha \sim
0.25$ (Sari et al. 1998); {\it ii)\/} If a spherical fireball slows 
down to a non-relativistic expansion (Dai and Lu 1999) then $\Delta
\alpha = (\alpha_1+3/5)= 1.09$ for our value of $\alpha_1$;
{\it iii)\/} If the outflow is collimated with a fixed opening angle, 
the break in the light curve occurs when the relativistic beaming of 
the synchrotron radiation becomes wider than the jet opening angle 
with a predicted steepening of $\Delta \alpha =3/4$
(M\'esz\'aros and Rees 1999); {\it iv)\/} finally, if the afterglow
arises in a sideways expanding jet, the steepening will be $\Delta
\alpha =(1-\alpha_1/3)=0.44$ (Rhoads 1999) for our value of 
$\alpha_1$. The above estimates all assume a constant mean density 
distribution of the ambient medium. Only model {\it iii}, i.e. a
jet with fixed opening angle, is consistent with the observed value
of $\Delta \alpha = 0.70\pm0.09$. This model predicts a spectral
slope of the afterglow of $\beta = 2 \alpha_1/3 = 1.13 \pm 0.01$,
which is consistent with the 
$\beta = 1.00 \pm 0.18$ from the multi-band photometry. 
If the density of the surrounding medium was that of stellar wind
($n \propto r^{-\delta}$ with $\delta = 2$) we expect 
$\Delta \alpha = \frac{3-\delta}{4-\delta} = 0.50$ (M\'esz\'aros and Rees
1999; Jaunsen et al. 2001), which is excluded by the data at the
2.2$\sigma$ level.

\subsection{Comparison with GRB~000301C}
  Even though the gamma-ray emission of GRB~000301C (Jensen et al. 2001) 
was of about 10 times shorter duration than that of GRB~000926, 
the fact that they have nearly identical redshifts
of z=2.0375$\pm$0.0007 for GRB~000926 (M\o ller et al. in prep)
and z=2.0404$\pm$0.0008 for GRB~000301C (Jensen et al. 2001) makes it 
very convenient to compare the two. Both GRB~000301C and GRB~000926
displayed broken power-law decays. For the OT of GRB~000301C
Jensen et al. (2001) determined $\beta = 0.70\pm0.09$ and 
$\Delta \alpha = 1.57\pm0.18$. In this case the best model is 
that of a side-ways expanding jet in a medium of constant density,
whereas a jet with fixed opening angle is not consistent with the 
data. Therefore, even though the two bursts appear similar they cannot
be explained by the same model.

  GRB~000301C and GRB~000926 have very different host galaxies.
The host galaxy of GRB~000301C remains undetected despite a very
deep detection limit of R=28.5 (Fruchter et al. 2000a; Smette et al.
2001), whereas the
host galaxy of GRB~000926 is relatively bright at R=23.87$\pm$0.15
(Sect. 6). Hence, the
host galaxy of GRB~000926 is more than 70 times brighter than that
of GRB~000301C. In the same way \object{GRB~990123} and \object{GRB~990510}
occured
at nearly identical redshifts (z$\approx$1.6) and the host galaxy 
of the former is 
more than 30 times brighter than the latter (Holland and Hjorth 1999;
Fruchter et al.
1999, 2000b). If GRBs indeed trace star-formation these observations
indicate that at these redshifts galaxies covering a broad range 
of luminosities contribute significantly to the over-all density of
star formation. Furthermore, as the observed R--band flux is 
proportional to the star formation rate, there must be 
1--2 orders of magnitude more galaxies at the R=28 level than at the 
R=24 level at z$\approx$2. Otherwise it would be unlikely to detect
R=28 galaxies as GRB hosts (under the assumption that GRBs trace
star-formation). An alternative explanation is that the
faint host galaxies of GRB~000301C and GRB~990510 are faint at
rest-frame UV wavelengths due to massive extinction similar to some
sources selected in the sub-mm range (e.g. Ivison et al. 2000). 
However, the low extinction derived from the optical properties of the 
GRB~000301C afterglow argues against this explanation at least for this
particular burst. 

\section*{Acknowledgements}
Most of the optical data presented here have been taken using ALFOSC, which
is owned by the Instituto de Astrofisica de Andalucia (IAA) and operated at
the Nordic Optical Telescope under agreement between IAA and the NBIfAFG of
the Astronomical Observatory of Copenhagen.  UKIRT is operated by the Joint
Astronomy Centre on  behalf of the  Particle Physics and Astronomy Research
Council of the United Kingdom. JUF and THD acknowledges enthusiastic help and
support from C. M{\o}ller and I. Sv{\"a}rdh during the hectic moments of
finding the OT  by comparison with  DSS-plates. JUF acknowledges H.O. Fynbo
for introducing him to CERNs MINUIT  fitting programme. JG acknowledges the
receipt of a  Marie Curie Research Grant from  the European Commission. MIA
acknowledges the Astrophysics group of the Physics dept. of
University of Oulu for support of his work. IRS acknowledges support
from a Royal Society URF. IB
was   supported   by    P\^ole   d'Attraction    Interuniversitaire,  P4/05
\protect{(SSTC, Belgium)}. JMCC acknowledges the receipt of a FPI
doctoral fellowship from Spain's Ministerio de Ciencia y Tecnolog{\'i}a.
KH is  grateful  for Ulysses support under  JPL
Contract 958056, and for NEAR support under NASA grants NAG5-9503 and
NAG5-3500.
Additionally, the availability of the GRB Coordinates Network (GCN) and
BACODINE services, maintained by Scott Barthelmy, is greatly
acknowledged.
We acknowledge the availability of POSS-II exposures, used in this work;
The Second Palomar Observatory Sky Survey (POSS-II) was made by the
California
Institute of Technology with funds from the National Science Foundation,
the National Aeronautics and Space Administration, the National
Geographic Society, the Sloan Foundation, the Samuel Oschin Foundation,
and the Eastman Kodak Corporation. We acknowledge the availability
of the 2MASS catalogs.
This work was supported by the Danish Natural Science Research Council (SNF).

\end{document}